\documentclass[aps,twocolumn,showpacs]{revtex4}
\begin{document}
\bibliographystyle{apsrev}
\def\half{{1\over 2}}
\def \D {\mbox{D}}
\def\curl {\mbox{curl}\,}
\def \ep {\varepsilon}
\def \lleq {\lower0.9ex\hbox{ $\buildrel < \over \sim$} ~}
\def \ggeq {\lower0.9ex\hbox{ $\buildrel > \over \sim$} ~}
\def\beq{\begin{equation}}
\def\eeq{\end{equation}}
\def\ber{\begin{eqnarray}}
\def\eer{\end{eqnarray}}
\def \apl {ApJ, }
\def \aps {ApJS, }
\def \pd {Phys. Rev. D, }
\def \prl {Phys. Rev. Lett., }
\def \pl {Phys. Lett., }
\def \np {Nucl. Phys., }
\def \l {\Lambda}

\textheight 22.0cm
\textwidth 16cm
\leftmargin -4cm
\topmargin -1cm

\title{IMPLEMENTING POWER LAW INFLATION WITH  TACHYON ROLLING ON THE BRANE}
\author{
M. Sami}
\email{sami@jamia.net}
\affiliation{Department of Physics, \\
Jamia Millia Islamia, New Delhi-110025, INDIA}
\begin{abstract}

We study a minimally coupled tachyon field rolling down to its ground
state on the FRW brane. We construct tacyonic potential which can
implements power law inflation in the brane world cosmology. The potential
turns out to be ${V_0 \phi^{-1}}$ on the brane and reduces  to inverse square potential
at late times when brane corrections to the Friedmann equation become
negligible. We also do similar exercise with a normal scalar field and
discover that the inverse square potential on the brane leads to power
law inflation.

\end{abstract}
\maketitle 

\section{Introduction}
 Cosmological inflation has become an integral part of the standard
 model of the universe. Apart from being
 capable of removing the shortcomings of the standard cosmology,
the paradigm seems to have gained a fairly good amount of
support from the recent observations on microwave background
radiation. On the other hand there have been difficulties in
obtaining accelerated expansion from fundamental theories such
as M/String theory. Recently, Sen  \cite{sen1,sen2} has shown that the decay
of an unstable D-brane produces pressure-less gas 
with finite energy density that resembles classical dust. Gibbons has emphasized
the cosmological implications of  tachyonic condensate rolling towards its ground state.
\cite{gibbons}. Rolling tachyon
matter associated with unstable D-branes has an interesting
equation of state which smoothly interpolates between -1 and
0. The tachyonic matter, therefore, might provide an explanation
for inflation at the early epochs and could contribute to some
new form of dark matter at late times \cite{ftytgat,feinstein,shiu,paddy2,paddy,jatkar,hashimoto,kim,shima,minahan,cornalba,benoum,li,hwang} . 
 An effective potential for tachyon
condensate is computed  in reference \cite{moore}; the expression
is exact in $\alpha'$ but tree level in $g_s$.
Sen \cite{sen3} has shown
that the choice of an exponential potential for the tachyonic
field leads to the absence of plane-wave solutions around the
tachyon vacuum and exponential decay of the pressure at late
times. However, due to the limitations of non-perturbative
calculation, it is perfectly legitimate, as pointed out
by Padmanabhan \cite{paddy}, to construct a potential leading to desired
cosmological evolution.The evolution of FRW cosmological models
and linear perturbations of tachyon
matter rolling towards a minimum of its potential is discussed
by   Frolov ,
 Kofman and  Starobinsky \cite{kstarobinsky}.\\

    Another interesting development in cosmology inspired by
String theory is related to Brane World cosmology.	In this
scenario, our four dimensional space time is realized as an
embedding or a boundary (brane) of the higher dimensional
space time (bulk). In this picture all the matter fields are
confined to the brane whereas gravity can propagate in the
bulk. The scenario has interesting cosmological implications,
in particular, the prospects of inflation are enhanced on the
brane due to an additional quadratic term in density in the
Friedmann equation. In this note , following reference \cite{paddy}
, we construct potentials which can implement power law
inflation on FRW brane in presence of normal scalar field
as well the tachyonic field. The fact that one can construct $ V(\phi)$ for a given $a(t)$ has been known
earlier \cite{ellis} and the method was practically used in reference \cite{sstarobinsky}.
Effects of tachyon in context
of brane world cosmology are discussed in reference \cite{papas}. Dynamics of gauge fields with rolling tachyon on unstable D-branes
is studied in reference \cite{minahan2,akira,mehen}

\subsection{ SCALAR FIELD POTENTIAL FOR NON-TACHYONIC MATTER ON
FRW BRANE} In the 4+1 dimensional brane world scenario inspired by
Randall-Sundrum model \cite{randall}, the Friedmann equation is modified to\cite{cline} 
\begin{equation} H^2={1 \over 3M_p^2} \rho \left(1+{\rho \over 2\lambda_b}
\right)+ {\Lambda_4 \over 3}+{ {\cal E} \over a^4} \end{equation} where $\cal E$ is
an integration constant which transmits bulk graviton influence onto
the brane and $\lambda_b$ is the brane tension.  For simplicity we set
$\Lambda_4$ equal to zero and also drop the last term as otherwise the
inflation would render it so, leading to the expression \begin{equation}
H^2={1 \over 3M_p^2} \rho \left(1+{\rho \over 2\lambda_b} \right)
\end{equation} 
where $ \rho \equiv \rho_{\phi}={1 \over 2} \dot{\phi}^2+V(\phi) $ if one is dealing with universe
 dominated by a single scalar field minimally coupled
to gravity. The
pressure of the scalar field is given by
 $p \equiv p_{\phi}={1 \over 2} \dot{\phi}^2-V(\phi)$ and the equation of motion of the field propagating on the
brane is \begin{equation} \ddot{\phi}+3H \dot{\phi}+{dV \over d{\phi}}=0
\end{equation}
	  The conservation equation equivalent to ( 3 ) is
\begin{equation} {\dot{\rho}_{\phi} \over \rho_{\phi}}+3H (1+\omega)=0
\end{equation} where $\omega={p_{\phi} \over \rho_{\phi}}$ is the equation
of state for the scalar field.	The brane effects in context of inflation
are most pronounced in the high energy limit $V>> \lambda_b$ ; the
Friedmann equation in this limit becomes 
\begin{equation} 
H={\rho \over{(
6\lambda_b M_p^2)^{1/2}}} 
\end{equation} 
Using the Friedmann equation (
5 ) and the conservation equation ( 4 ), one obtains the expression for
$(1+\omega)$ \begin{equation} 1+\omega=-{1 \over 3} {\dot{H} \over H^2}
\end{equation} From the evolution equation ( 3 ) and the expression $
{\dot{\phi}^2 \over 2}=V{(1+\omega) ( 1 - \omega)^{-1}}$, one obtains the
differential equation for the potential V \begin{equation} {\dot{V} \over
V}=-{{\dot{f}+6Hf} \over {1+f}} \end{equation} where $ f=(1+w)(1-w)^{-1}$
Integrating ( 7 ) leads to the expression for V as a function of time

\begin{equation}
V(t)={{C} \over {6 H}}\left( \dot{H} +6H^2\right)
\end{equation}
where $C=(6 \lambda_b M_p^2)^{1/2}$. Expressing $f(t)$ in terms of H and its derivative and using equation ( 8 ) leads to
the expression of $\phi(t)$
\begin{equation}
\phi(t)=\int{\left(-{C \over 3} {{\dot{H}} \over H}\right)^{1/2}}dt
\end{equation}
Equations (8) and ( 9 ) can be used to determine $\phi(t)$ and $V(t)$ for a given a(t) on the brane.  
For $ a(t) \propto t^n$
\begin{equation}
V(t)=C\left(1-{1\over 6n}\right){n \over t}
\end{equation}
\begin{equation}
\phi(t)-\phi_0=\sqrt{4 C \over 3}t^{1/2}
\end{equation}
. Combining ( 10 ) and ( 11 ) we obtain the expression for the potential as a function of $\phi$
\begin{equation}
V(\phi)={\lambda_b \over 2} \left(1-{1\over 6n}\right){n \over \left( {{\phi(t)-\phi_0} \over M_p}\right)^2}
\end{equation}
where $n\ > 1/6$.
In usual 4-dimensional FRW cosmology, the power law inflation is implemented by an exponential potential whereas its counter
part on the brane turns out to be very different. Indeed, we earlier investigated the prospects of inflation with
 exponential potential on the brane, in a different context, and found the exact solution of the problem. In particular, a(t) was shown to have
 complicated time dependence through double exponential\cite{ssami} .
\subsection{POTENTIAL FOR TACHYON ROLLING ON THE BRANE}
As recently demonstrated by Sen \cite{sen1,sen2}, a rolling tacyhon condensate in a specially flat FRW cosmological model is described
by an effective fluid with energy momentum tensor $T^{\mu}_{\nu}=diag\left(-\rho,p,p,p\right)$, where the energy density
$\rho$ and pressure p are given by
\begin{equation}
\rho={V(\phi) \over {\sqrt{1-\dot{\phi}^2}}}
\end{equation}
\begin{equation}
p=-V(\phi)\sqrt{1-\dot{\phi}^2}
\end{equation}
The Friedmann equation on the FRW brane($V(\phi)>>2\lambda_b$) for the tachyonic matter
 is same as equation ( 5 ) with energy density given by ( 13 ).
For the tachyonic field the expression for $(1+\omega)$ is also given by equation ( 6 ); however, $\omega$ in the present
case has a form
\begin{equation}
\omega=\dot{\phi}^2(t)-1
\end{equation}
Making use of equations (6), (13), (14) and (15) we get the expressions for $V(t)$ and $\phi(t)$
\begin{equation}
\phi(t)=\int{\left(-{{\dot{H} \over {3H^2}}}\right)^{1/2}} dt
\end{equation}
\begin{equation}
V(t)=C H\left(1+{\dot{H} \over 3H^2}\right)^{1/2}
\end{equation}
Equations (16) and ( 17 ) analogous to (9) and (10) determine the tachyonic field and the tachyonic potential
as a function of time for a given scale factor a(t). In case of power law expansion of the universe $ a(t) \propto t^n$ we obtain
\begin{equation}
\phi(t)-\phi_0=\left({1 \over 3n}\right)^{1/2}t
\end{equation}
\begin{equation}
V(t)= C \left(1-{1 \over 3n}\right)^{1/2}{n \over t}
\end{equation}
Using Eqs. (19) and (18) we obtain the expression for $V(\phi)$
\begin{equation}
V(\phi)=2\lambda_b \sqrt{n}\left(1-{1\over 3n}\right)^{1/2}{1 \over \left( {{\phi(t)-\phi_0} \over M_p}\right)}
\end{equation}
where $ n\ > 1/3 $. Taking the large value of n one may have the desired accelerated expansion with $V(\phi) $ given by
( 20 ). However, a comment regarding the range of n is in order. There are string theory arguments that the tachyonic matter asymptotically
evolves to a pressure-less gas , i.e. $\dot{\phi}(\infty)=1$ irrespective of the form of  tachyonic potential. And this in tern implies that at late times
$a(t) \propto t^{2/3}$. In fact, the expressions obtained here are valid in the limit $ V>> \lambda_b$ which is 
a legitimate limit at times inflation was operative. But at late times the tachyon field rolls down its potential
to the extent that $ V<< \lambda_b$ and in this limit the brane corrections disappear and equation ( 2 ) reduces
to usual Friedmann equation allowing the scale factor to evolve asymptotically as 
$a(t) \propto t^n$ with $n=2/3$ [ 7 ].\par
 To sum up, we have constructed potentials on the FRW brane with and without tachyonic matter 
 which successfully implement power law inflation. It would be interesting to carry out phase space analysis 
 with tachyonic potential to investigate the late time behavior of the power law inflation in usual FRW cosmology
 as well as in the brane world scenario.\\   
Inspite of the very attractive features of the rolling tachyon condensate, the tachyonic inflation
faces difficulties associated with reheating \cite{klinde,sami}. A homogeneous tachyon field evolves towards its
ground state without oscillating about it and , therefore, the  conventional reheating
mechanism in tachyonic model does not work. Quantum mechanical particle
production during inflation  provides an alternative mechanism by means of
which the universe could reheat. Unfortunately, this mechanism also does not seem to
work: the  small energy density of
radiation  created in this process red-shifts faster than the energy density of the tachyon field. However, the tachyon field
could play the dual role of quintessence and dark matter \cite{paddy2}.
\begin{acknowledgements}
I am thankful to V. Sahni, A. Starobinsky and V. Johri for useful discussions.
\end{acknowledgements}

\end{document}